\documentclass[a4paper]{jpconf}

\usepackage{graphicx}

\usepackage{amsmath, amssymb, amsfonts}
\usepackage{braket}
\usepackage{color}

\newcommand{\kket}[1]{| #1 \rangle\!\rangle}

\newcommand{\bbrakket}[1]{\langle\!\langle #1 \rangle\!\rangle}

\usepackage{textpos}
\usepackage{etoolbox}
\patchcmd{\thebibliography}{\advance\leftmargin\labelsep}
  {\labelsep=0.5cm \advance\leftmargin\labelsep}{}{}
  
\begin{document}

\title{Influence functionals, decoherence and\\ conformally coupled scalars }

\author{C Burrage$^1$, C K\"{a}ding$^1$, \underline{P Millington}$^1$ and  J Min\'{a}\v{r}$^{2,3,1,4}$}

\address{$^1$ School of Physics and Astronomy, University of Nottingham, \\ Nottingham NG7 2RD, UK}
\address{$^2$ Institute for Theoretical Physics, University of Amsterdam, \\ Science Park 904, 1098 XH Amsterdam}
\address{$^3$ Department of Physics, Lancaster University,\\ Lancaster, LA1 4YB, UK}
\address{$^4$ Centre for the Mathematics and Theoretical Physics of Quantum Non-Equilibrium Systems, School of Physics and Astronomy, University of Nottingham, Nottingham, NG7 2RD, UK}

\ead{clare.burrage@nottingham.ac.uk, christian.kading@nottingham.ac.uk, p.millington@nottingham.ac.uk, jminar@uva.nl}

\begin{textblock}{4}(12,-9.2)
\begin{flushright}
\begin{footnotesize}
11 October 2019
\end{footnotesize}
\end{flushright}
\end{textblock}

\begin{abstract}
Some of the simplest modifications to general relativity involve the coupling of additional scalar fields to the scalar curvature. By making a Weyl rescaling of the metric, these theories can be mapped to Einstein gravity with the additional scalar fields instead being coupled universally to matter. The resulting couplings to matter give rise to scalar fifth forces, which can evade the stringent constraints from local tests of gravity by means of so-called screening mechanisms. In this talk, we derive evolution equations for the matrix elements of the reduced density operator of a toy matter sector by means of the Feynman-Vernon influence functional. In particular, we employ a novel approach akin to the LSZ reduction more familiar to scattering-matrix theory. The resulting equations allow the analysis, for instance, of decoherence induced in atom-interferometry experiments by these classes of modified theories of gravity.
\end{abstract}

\section{Introduction}

There are many pointers to a need to extend the Standard Model of particle physics (SM) and/or general relativity, not least the dark matter and dark energy problems, and the simplest extensions often involve additional scalar fields. If these new fields are neutral under the SM gauge groups, they may couple to the SM via the Higgs portal or through a non-minimal coupling to the Ricci scalar (of Brans-Dicke type~\cite{Brans:1961sx}), mediating interactions with hidden sectors. In fact, for the SM, Higgs-portal and non-minimal gravitational couplings are related to one another (up to and including dimension-four operators) by a change of conformal frame~\cite{Burrage:2018dvt}.

If the scalar fields are heavy on astrophysical scales and stable on cosmological time-scales, they may compose the relic density of dark matter, produced thermally in the early universe. However, with the absence of a direct or indirect detection of a particulate dark matter candidate, interest has been growing in new light fields, which can yield additional long-range forces relevant on astrophysical scales~\cite{Burrage:2016yjm, OHare:2018ayv, Burrage:2018zuj} and be constrained by terrestrial atom-interferometry experiments~\cite{Burrage:2014oza, Hamilton:2015zga, Schlogel:2015uea, Elder:2016yxm, Jaffe:2016fsh, Brax:2016wjk, Brax:2017xho}. Cosmologically light fields may also play a role in the late-time accelerated expansion, where, e.g., quintessence scenarios of dark energy can be realized through the dynamics of a pseudo-Goldstone boson of a spontaneously broken scale symmetry~\cite{GarciaBellido:2011de}.

Any additional scalar forces operative on Solar System scales are heavily constrained by local tests of gravity, as well as terrestrial metrology experiments. In order to evade these constraints, and without fine-tuning couplings to matter, we can appeal to so-called screening mechanisms, wherein the fifth force is suppressed dynamically in regions of high ambient density. Around a point source of unit mass, the scalar perturbations $\delta X$ above some (approximately) constant background $\bar{X}$ satisfy an equation of motion schematically of the form~\cite{Joyce:2014kja}
\begin{equation}
-\:Z(\bar{X})\big(\ddot{\delta X}-c_s^2(\bar{X})\nabla^2\delta X\big)\:-\:M^2(\bar{X})\delta X\ =\ \frac{{\rm d}A(\bar{X})}{{\rm d}\bar{X}}\,\delta(\mathbf{x})\;,
\end{equation}
giving rise to a Yukawa potential
\begin{equation}
U(r)\ \sim \ -\:\frac{1}{Z(\bar{X})c_s^2(\bar{X})}\bigg[\frac{{\rm d} A(\bar{X})}{{\rm d}\bar{X}}\bigg]^2\frac{1}{4\pi r}\exp\bigg[-\:\frac{M(\bar{X})\,r}{Z^{1/2}(\bar{X})c_s(\bar{X})}\bigg]\;.
\end{equation}
The fifth force can therefore be suppressed by dynamically modifying the mass $M(\bar{X})$, as in the chameleon mechanism~\cite{Khoury:2003aq, Khoury:2003rn}; the coupling function $A(\bar{X})$, as in the symmetron mechanism~\cite{Hinterbichler:2010es, Hinterbichler:2011ca} (for related models, see references~\cite{Dehnen:1992rr, Gessner:1992flm, Damour:1994zq, Pietroni:2005pv, Olive:2007aj, Brax:2010gi, Burrage:2016xzz}), or the kinetic structure $Z(\bar{X})c_s^2(\bar{X})$, as in the Vainshtein~\cite{Vainshtein:1972sx} mechanism (see also references~\cite{Nicolis:2004qq, Nicolis:2008in, Babichev:2009ee, Brax:2012jr, Burrage:2014uwa}).

In this note, we consider the impact of an environment composed of a light, screened scalar field on a probe system, reporting the results of reference~\cite{Burrage:2018pyg}. Proceeding from first principles by means of the Feynman-Vernon influence functional~\cite{Feynman:1963fq} (see also reference~\cite{Calzetta_1998}), we derive the quantum master equation for the reduced density matrix of a toy `atomic' system (having in mind an atom-interferometry experiment). Assuming that `multi-atom' states are suppressed, we project into the single-particle subspace by means of an LSZ-like reduction technique~\cite{Lehmann:1954rq}, before commenting on how the renormalization procedure must be adapted to account for finite-time effects. While the impact of the screened scalar is beyond the reach of current experiments, as one might expect, screened scalar fields nevertheless provide a solid context within which to develop this general and robust approach to studying decoherence in quantum field theory.

\section{Conformally coupled scalars}
\label{sec:model}

The action of the models that we have in mind can be written in the general form
\begin{align}
S\ &=\ \int_x\;\sqrt{-\,g}\;\bigg[\frac{1}{2}\,M_{\rm Pl}^2\mathcal{R}\:-\:\frac{1}{2}\,g^{\mu\nu}\,\partial_{\mu}X\,\partial_{\nu}X\:-\:V(X)\bigg]\:+\:\int_x\;\sqrt{-\,\tilde{g}}\;\tilde{\mathcal{L}}_m(\{\tilde{\phi}_i\},\tilde{g}_{\mu\nu})\;,
\end{align}
where $M_{\rm Pl}$ is the reduced Planck mass, $\mathcal{R}$ is the Ricci scalar with respect to the Einstein-frame metric $g_{\mu\nu}$ (see reference \cite{FujiiMaeda}) and $\tilde{\mathcal{L}}_m$ is the matter Lagrangian. The matter fields $\{\tilde{\phi}_i\}$ move on geodesics of the Jordan-frame metric $\tilde{g}_{\mu\nu}=A^2(X)g_{\mu\nu}$, coupling universally to the scalar field $X$. We employ the $(-,+,+,+)$ signature convention and use the shorthand notation \smash{$\int_x\equiv \int{\rm d}^4x$}.

As a proxy for our `atom', we take a massive scalar field theory with Lagrangian
\begin{equation}
\tilde{\mathcal{L}}_m\ =\ -\:\frac{1}{2}\,\tilde{g}^{\mu\nu}\,\partial_{\mu}\tilde{\phi}\,\partial_{\nu}\tilde{\phi}\:-\:\frac{1}{2}\,\tilde{m}^2\,\tilde{\phi}^2\;.
\end{equation}
The matter action then has the following form in terms of the Einstein-frame metric:
\begin{equation}
S_m\ =\ \int_x\;\sqrt{-\,g}\;\bigg[-\:\frac{1}{2}\,A^2(X)\,g^{\mu\nu}\,\partial_{\mu}\tilde{\phi}\,\partial_{\nu}\tilde{\phi}\:-\:\frac{1}{2}\,A^4(X)\,\tilde{m}^2\tilde{\phi}^2\bigg]\;.
\end{equation}
After defining $\phi \equiv A(X)\tilde{\phi}$, so as to obtain a near-canonical kinetic term, we have
\begin{align}
S_m\ &=\ \int_x\;\sqrt{-\,g}\;\bigg[-\:\frac{1}{2}\,g^{\mu\nu}\,\partial_{\mu}\phi\,\partial_{\nu}\phi\:-\:\frac{1}{2}\,\phi^2\,g^{\mu\nu}\,\partial_{\mu}\ln A(X)\,\partial_{\nu}\ln A(X)\nonumber\\&\qquad+\:\phi\,g^{\mu\nu}\,\partial_{\mu}\phi\,\partial_{\nu}\ln A(X)\:-\:\frac{1}{2}\,A^2(X)\,\tilde{m}^2\phi^2\bigg]\;.
\end{align}
The equation of motion for the classical expectation value of the universally coupled scalar $\braket{X}$~is
\begin{equation}
\Box \braket{X}\:-\:\frac{{\rm d}V(X)}{{\rm d} X}\bigg|_{X\,=\,\braket{X}}\ =\ -\:\frac{1}{2}\,\frac{{\rm d}A^2(X)}{{\rm d}X}\bigg|_{X\,=\,\braket{X}}\tilde{T}\;,
\end{equation}
where, in our case, $\tilde{T}$ is the trace of the energy-momentum tensor of the vacuum chamber of the atom-interferometry experiment, taking the form $\tilde{T}=-A^{-1}(\braket{X})\rho^{\rm ext}$, where $\rho^{\rm ext}$ is the covariantly conserved energy density (see, e.g., reference~\cite{Joyce:2014kja}).

We now specialize to the quartic chameleon~\cite{Gubser:2004uf} with potential and coupling function
\begin{equation}
V(X)\ =\ \frac{\lambda}{4!}\,X^4\qquad \text{and}\qquad A(X)\ =\ e^{X/\mathcal{M}}\;.
\end{equation}
While the parameter space for this model is heavily constrained by atom-interferometry and torsion-balance experiments~\cite{Burrage:2017qrf}, it provides a convenient context for our discussions.

Working to leading order in the expansion of the coupling function, and in a constant density environment, the chameleon field acquires a background value of
\begin{equation}
\braket{X}\ =\ -\bigg(\frac{6\rho^{\rm ext}}{\lambda \mathcal{M}}\bigg)^{\!1/3}\;,
\end{equation}
with the fluctuations $\chi\equiv X-\braket{X}$ about the minimum of the effective potential $V^{\rm eff}=V+\rho^{\rm ext}X/\mathcal{M}$ having a density-dependent mass
\begin{equation}
M^2\ =\ \bigg(\frac{\lambda}{2}\bigg)^{\!1/3}\bigg(\frac{3\rho^{\rm ext}}{\mathcal{M}}\bigg)^{\!2/3}\;,
\end{equation}
which vanishes in vacuum. Hence, in regions of high density, the mass of the fluctuations becomes large, and the Yukawa potential induced by the chameleon field is suppressed.

Working with operators up to dimension four only, the action for the `atom' field $\phi$ and the chameleon fluctuations $\chi$ can be separated into free and (self-)interacting parts as
\begin{subequations}
\label{eq:actions}
\begin{align}
S_\phi[\phi] \ &=\ \int_x \bigg[-\:\frac{1}{2}\,g^{\mu\nu}\,\partial_{\mu}\phi\,\partial_{\nu}\phi\: -\: \frac{1}{2}\,m^2 \phi^2\bigg]\;,
\\
S_\chi[\chi] \ &=\ \int_x  \bigg[-\:\frac{1}{2}\,g^{\mu\nu}\,\partial_{\mu}\chi\,\partial_{\nu}\chi\: -\:\frac{1}{2}\, M^2 \chi^2\bigg]\;,\\
S_{\chi,{\rm int}}[\chi] \ &=\  \int_{x\,\in\,\Omega_t}\bigg[-\:\frac{\lambda}{4!}\,\big(\chi^4 + 4\braket{X}\chi^3\big)\bigg]\;,\\
S_{\rm int}[\phi,\chi] \ &=\   \int_{x\,\in\,\Omega_t}
\bigg[-\:\frac{1}{2}\,\bigg(2\,\frac{m^2}{\mathcal{M}}\bigg) \chi\phi^2\:-\:\frac{1}{4}\,\bigg(4\,\frac{m^2}{\mathcal{M}^2}\bigg) \chi^2\phi^2\bigg]\;,
\end{align}
\end{subequations}
where we have defined (assuming $\braket{X}$ is constant)
\begin{equation}
\label{eq:massredef}
m^2\ \equiv\ \tilde{m}^2\bigg(1\:+\:2\frac{\braket{X}}{\mathcal{M}}\:+\:2\,\frac{\braket{X}^2}{\mathcal{M}^2}\bigg)\;.
\end{equation}
Our aim in the next two sections is to trace out the chameleon fluctuations and derive the quantum master equation describing the evolution of the density matrix of the remaining open scalar system. With this in mind, and in order to account for the fact that the initial state of the `atom' system will be prepared and its final state measured over a finite interval of time, we have restricted the support of the spacetime integrals in the interaction parts to the hypervolume $\Omega_t=[0,t]\times \mathbb{R}^3$ (see references~\cite{Millington:2012pf, Millington:2013isa}), approximating the vacuum chamber as infinitely large. Note that this amounts to an instantaneous switching on and off of the interactions.

\section{Open dynamics and the Feynman-Vernon influence functional}

Our starting point is the reduced density functional --- the matrix element of the density operator in the basis of field eigenstates --- of the `atom' system (presented here in a simplified notation, cf.~the full time-dependence and functional delta functions that appear in reference~\cite{Burrage:2018pyg})\vspace{-0.1em}
\begin{equation}
\rho_{\phi}[\phi^{\pm};t_f]\ =\ \int{\rm d}\chi_f^{\pm}\;\rho[\phi^{\pm},\chi^{\pm};t_f]\;,\vspace{-0.1em}
\end{equation}
where \smash{$\int{\rm d}\chi_f^{\pm} \equiv \int{\rm d}\chi_f^+\int{\rm d}\chi_f^-$} is an integral over the doubled field configurations at time $t_f$. This doubling of degrees of freedom necessarily arises when constructing a path-integral representation of a trace, leading to the  Schwinger-Keldysh closed-time-path (CTP) formalism~\cite{Schwinger:1960qe, Keldysh:1964ud}.

Assuming that the initial state of the system (at time $t_i$) is factorizable, i.e.\vspace{-0.1em}
\begin{equation}
\rho[\phi^{\pm},\chi^{\pm};t_i]\ =\ \rho_{\phi}[\phi^{\pm};t_i]\rho_{\chi}[\chi^{\pm};t_i]\;,\vspace{-0.1em}
\end{equation}
the reduced density functional evolves as\vspace{-0.1em}
\begin{equation}
\label{eq:rhoevofunc}
\rho_{\phi}[\phi^{\pm};t_f]\ = \int {\rm d}\phi_{i}^{\pm}\;\mathcal{I}[\phi^{\pm};t_f,t_i]\rho_{\phi}[\phi^{\pm};t_i]\;,\vspace{-0.1em}
\end{equation}
where
\begin{equation}\vspace{-0.1em}
\mathcal{I}[\phi^{\pm};t_f,t_i]\ =\ \int_{\phi_i^{\pm}}^{\phi_f^{\pm}}\mathcal{D}\phi^{\pm}\;e^{i\widehat{S}_{\rm eff}[\phi;t_f,t_i]}\vspace{-0.1em}
\end{equation}
is the influence-functional propagator ($\hbar=1$). The effective action $\widehat{S}_{\rm eff}[\phi;t_f,t_i]$ can be written
\begin{equation}\vspace{-0.1em}
\widehat{S}_{\rm eff}[\phi;t_f,t_i]\ = \ \widehat{S}_{\phi}[\phi]\:+\:\widehat{S}_{\rm IF}[\phi;t_f,t_i]\;.\vspace{-0.1em}
\end{equation}
We use a $\widehat{\ }$ to identify actions that depend on both of the doubled degrees of freedom. Doubled actions corresponding to unitary evolution, i.e.~obtained from equation~\eqref{eq:actions}, have, e.g., a form
\begin{equation}\vspace{-0.1em}
\widehat{S}_{\phi}[\phi]\ =\ S_{\phi}[\phi^+]\:-\:S_{\phi}[\phi^-]\;.\vspace{-0.1em}
\end{equation}
The influence action $\widehat{S}_{\rm IF}[\phi;t_f,t_i]$ on the other hand, which arises after tracing out the $\chi$ field, mixes the $+$ and $-$ type fields. It is defined via the Feynman-Vernon influence functional\vspace{-0.1em}
\begin{align}
\mathcal{F}[\phi^{\pm}]\ &=\  e^{i\widehat{S}_{\rm IF}[\phi]}\ = \ \int{\rm d}\chi_f^{\pm}\int{\rm d}\chi_i^{\pm}\;\rho_{\chi}[\chi^{\pm};t_i]\int_{\chi_i^{\pm}}^{\chi_f^{\pm}}\mathcal{D}\chi^{\pm}\;e^{i(\widehat{S}_{\chi}[\chi]+\widehat{S}_{\chi,{\rm int}}[\chi]+\widehat{S}_{\rm int}[\phi,\chi])}\;,\vspace{-0.1em}
\end{align}
wherein we have omitted time arguments for brevity. Taking the partial time derivative of the reduced density functional in equation~\eqref{eq:rhoevofunc}, we obtain the functional expression of the master equation (with $t\equiv t_f$). If the influence action contains only local interactions, we obtain\vspace{-0.1em}
\begin{equation}
\partial_t\rho_{\phi}[\phi^{\pm};t]\ =\ -\:i\big(-\partial_t S_{\phi}[\phi^+]+\partial_tS_{\phi}[\phi^-]\big)\rho_{\phi}[\phi^{\pm};t]\:+\:i\partial_t\widehat{S}_{\rm IF}[\phi;t]\rho_{\phi}[\phi^{\pm};t]\;,\vspace{-0.1em}
\end{equation}
wherein we can identify the Hamiltonian (first two terms) and Lindblad terms (third term).

\section{LSZ-like reduction}

The path-integral approach detailed in the previous section has the advantage that we can make use of the powerful and familiar diagrammatic techniques to evaluate the influence functional. However, we would like to work in a basis other than that of the field eigenstates. As in the case of scattering-matrix theory, we can project into the Fock space, and we will limit ourselves here to the single-particle subspace, by means of an LSZ-like reduction technique~\cite{Lehmann:1954rq}.

In order to motivate the form of this reduction, it is convenient to work in the canonical operator approach known as thermo field dynamics~\cite{Takahasi:1974zn, Arimitsu:1985ez, Arimitsu:1985xm}. Therein, we work with a doubled Hilbert space $\widehat{\mathcal{H}}=\mathcal{H}^+\otimes \mathcal{H}^-$ (cf.~the notation of reference~\cite{Khanna}), which allows the quantum Liouville equation (in the Schr\"{o}dinger picture) to be recast in a Schr\"{o}dinger-like form
\begin{equation}
\partial_t\kket{\rho(t)}\ =\ -\,i\widehat{H}\kket{\rho(t)}\;.
\end{equation}
Herein, $\widehat{H}$ is the Liouvillian operator, which, for an isolated system, takes the form
\begin{equation}
\widehat{H}\ =\ \hat{H}\otimes \hat{\mathbb{I}}\:-\:\hat{\mathbb{I}}\otimes \hat{H}\;,
\end{equation}
where $\hat{H}$ is the Hamiltonian and $\hat{\mathbb{I}}$ is the identity operator. The density operator $\hat{\rho}(t)$ is replaced by a state in the doubled Hilbert space, taking the form
\begin{equation}
\kket{\rho(t)}\ =\ \hat{\rho}^+(t)\kket{1}\ \equiv\ \big(\hat{\rho}(t)\otimes \hat{\mathbb{I}}\big)\kket{1}\;.
\end{equation}
The state $\kket{1}$ can be written schematically as 
\begin{equation}
\kket{1}\ =\  \sum_n\ket{n^+}\otimes \ket{n^-}\;,
\end{equation}
where $n$ labels a complete basis. The expectation value of an operator $\hat{O}(t)$ is then
\begin{equation}
\braket{\hat{O}(t)}\ =\ {\rm tr}\,\hat{O}(t)\hat{\rho}(t)\ =\ \bbrakket{1|\big(\hat{O}(t)\otimes \hat{\mathbb{I}}\big)\big(\hat{\rho}(t)\otimes \hat{\mathbb{I}}\big)|1}\ =\ \bbrakket{1|\hat{O}^+(t)|\rho(t)}\;.
\end{equation}

Working in the interaction picture, we are interested in the single-particle matrix element
\begin{equation}
\label{eq:melemdef}
\rho(\mathbf{p},\mathbf{p}';t)\ = \ \braket{\mathbf{p};t|\hat{\rho}(t)|\mathbf{p}';t}\ =\ 	\mathrm{tr}\ket{\mathbf{p}';t}\!\bra{\mathbf{p};t}\hat{\rho}(t)\ =\ \bbrakket{1(t)|\big(\ket{\mathbf{p}';t}\!\bra{\mathbf{p};t}\otimes \hat{\mathbb{I}}\big)\hat{\rho}^+(t)|1(t)}
\end{equation}
in the basis of momentum eigenstates $\ket{\mathbf{p};t}$. Note that the states and operators are evaluated at equal times, ensuring that this matrix element is picture independent (see references~\cite{Millington:2012pf, Millington:2013isa}). Taking the time derivative, recalling that the state and basis vectors evolve respectively with the interaction and free parts of the effective Liouvillian operator $\widehat{H}_{\rm eff}$, we obtain 
\begin{align}
\label{eq:melemdot}
\partial_t\rho(\mathbf{p},\mathbf{p}';t)\ &=\ -\:i\int{\rm d}\Pi^{\phi}_{\mathbf{k}}\,{\rm d}\Pi^{\phi}_{\mathbf{k}'}\,\rho(\mathbf{k},\mathbf{k}';t)\bbrakket{0|{\rm T}[\hat{a}^+_{\mathbf{p}}(t)\hat{a}_{\mathbf{p}'}^-(t)\widehat{H}_{\rm eff}(t)\hat{a}^{+\dagger}_{\mathbf{k}}(t)\hat{a}_{\mathbf{k}'}^{-\dagger}(t)|0}\;,
\end{align}
where \smash{$E^{\phi}_{\mathbf{k}}=\sqrt{\mathbf{k}^2+m^2}$} and we use the shorthand notation
\begin{equation}
\int{\rm d}\Pi^{\phi}_{\mathbf{k}}\ \equiv\ \int_{\mathbf{k}}\frac{1}{2E^{\phi}_{\mathbf{k}}}\ =\ \int\!\frac{{\rm d}^3\mathbf{k}}{(2\pi)^32E^{\phi}_{\mathbf{k}}}
\end{equation}
for the Lorentz-invariant phase-space integral. By using
\begin{equation}
\hat{a}^+_{\mathbf{p}}(t)\ =\ +\,i\int_{\mathbf{x}}\;e^{-i\mathbf{p}\cdot\mathbf{x}}\,\partial_{t,E^{\phi}_{\mathbf{p}}}\hat{\phi}^+(t,\mathbf{x})\;,\qquad
\hat{a}^-_{\mathbf{p}}(t)\ =\ -\,i\int_{\mathbf{x}}\;e^{+i\mathbf{p}\cdot\mathbf{x}}\,\partial_{t,E^{\phi}_{\mathbf{p}}}^*\hat{\phi}^-(t,\mathbf{x})\;,
\end{equation}
where \smash{$\int_{\mathbf{x}}\equiv \int{\rm d}^3\mathbf{x}$} and 
\begin{equation}
\partial_{t,E^{\phi}_{\mathbf{p}}}\ \equiv\ \overset{\rightarrow}{\partial}_t\:-\:iE^{\phi}_{\mathbf{p}}\;,
\end{equation}
equation~\eqref{eq:melemdot} can be rewritten entirely in terms of field operators as
\begin{align}
\partial_t\rho(\mathbf{p},\mathbf{p}';t)\ &=\ -\:i\lim_{\substack{x^{0(\prime)}\,\to\, t^+\\y^{0(\prime)}\,\to\, 0^-}}\int{\rm d}\Pi^{\phi}_{\mathbf{k}}\,{\rm d}\Pi^{\phi}_{\mathbf{k}'}\;e^{i(E_{\mathbf{k}}^{\phi}-E_{\mathbf{k}'}^{\phi})t}\rho(\mathbf{k},\mathbf{k}';t)\int_{\mathbf{x}\mathbf{x}'\mathbf{y}\mathbf{y'}}e^{-i(\mathbf{p}\cdot\mathbf{x}-\mathbf{p}'\cdot\mathbf{x}')+i(\mathbf{k}\cdot\mathbf{y}-\mathbf{k}'\cdot\mathbf{y}')}\nonumber\\&\qquad\times\:\partial_{x^0,E^{\phi}_{\mathbf{p}}}\partial_{x^{0\prime},E^{\phi}_{\mathbf{p}'}}^*\partial_{y^0,E^{\phi}_{\mathbf{k}}}^*\partial_{y^{0\prime},E^{\phi}_{\mathbf{k}'}}\bbrakket{0|{\rm T}[\hat{\phi}^+(x)\hat{\phi}^-(x')\widehat{H}_{\rm eff}(t)\hat{\phi}^{+}(y)\hat{\phi}^{-}(y')]|0}\;,
\end{align}
where $x^{0(\prime)}\to t$ from above and $y^{0(\prime)} \to 0$ from below so that the operator time-ordering reproduces that in equation~\eqref{eq:melemdot}. This can be expressed in the path-integral language as
\begin{align}
\label{eq:startmaster}
&\partial_t\rho(\mathbf{p},\mathbf{p}';t)\ =\ i\lim_{\substack{x^{0(\prime)}\,\to\, t^+\\y^{0(\prime)}\,\to\, 0^-}}\int{\rm d}\Pi^{\phi}_{\mathbf{k}}\,{\rm d}\Pi^{\phi}_{\mathbf{k}'}\;e^{i(E_{\mathbf{k}}^{\phi}-E_{\mathbf{k}'}^{\phi})t}\rho(\mathbf{k},\mathbf{k}';t)\int_{\mathbf{x}\mathbf{x}'\mathbf{y}\mathbf{y'}}e^{-i(\mathbf{p}\cdot\mathbf{x}-\mathbf{p}'\cdot\mathbf{x}')+i(\mathbf{k}\cdot\mathbf{y}-\mathbf{k}'\cdot\mathbf{y}')}\nonumber\\&\quad\times\:\partial_{x^0,E^{\phi}_{\mathbf{p}}}\partial_{x^{0\prime},E^{\phi}_{\mathbf{p}'}}^*\partial_{y^0,E^{\phi}_{\mathbf{k}}}^*\partial_{y^{0\prime},E^{\phi}_{\mathbf{k}'}}\int\mathcal{D}\phi^{\pm}\;e^{i\widehat{S}_{\phi}[\phi]}\phi^+(x)\phi^-(x')\Big(\partial_t\widehat{S}_{\rm eff}[\phi;t]\Big)\phi^{+}(y)\phi^{-}(y')\;,
\end{align}
where we have replaced $\widehat{H}_{\rm eff}\to -\,\partial_t\widehat{S}_{\rm eff}$. Up to factors of the wavefunction normalization, the right-hand side contains the LSZ reduction~\cite{Lehmann:1954rq} of a CTP four-point correlation function.

\section{Quantum master equation}

Inserting the explicit form of the influence-functional action for the model of section~\ref{sec:model} (see reference~\cite{Burrage:2018pyg}) into equation~\eqref{eq:startmaster}, we obtain the master equation
\begin{align}
\label{eq:master}
\partial_t\rho(\mathbf{p},\mathbf{p}';t) \:  &=\: -\,i\big(E_{\mathbf{p}}^{\phi}-E_{\mathbf{p}'}^{\phi}\big)\,\rho(\mathbf{p},\mathbf{p}';t) \,-
\,\frac{im^2}{\mathcal{M}}\left(\frac{1}{E^{\phi}_{\mathbf{p}}}-\frac{1}{E^{\phi}_{\mathbf{p}'}}\right)\rho(\mathbf{p},\mathbf{p}';t) 
\nonumber
\\
&\quad\quad
\times \bigg\{
\frac{\Delta^{\rm F}_{xx}}{\mathcal{M}}
+
\bigg[\frac{m^2}{\mathcal{M}}\,D^{\rm F}_{xx}+
\frac{\lambda}{2}\braket{X}\Delta^{\rm F}_{xx}\bigg]\frac{\cos(Mt)-1}{M^2}\bigg\}
\nonumber
\\
&\phantom{=\ }
+\,i\,
\frac{4m^4}{\mathcal{M}^2}\sum_{s\,=\,\pm}\int_{\mathbf{k}}\Bigg\{\Bigg[
\rho(\mathbf{p},\mathbf{p}';t)\,\frac{1}{E^{\phi}_{\mathbf{p}}2E^{\chi}_{\mathbf{k}}2E^{\phi}_{\mathbf{p}-\mathbf{k}}}\,\frac{s}{\big(sE_{\mathbf{k}}^{\chi}+E^{\phi}_{\mathbf{p}-\mathbf{k}}\big)^2-\big(E_{\mathbf{p}}^{\phi}\big)^2}\nonumber\\&\qquad\times\Big[\Big(sE_{\mathbf{k}}^{\chi}+E^{\phi}_{\mathbf{p}-\mathbf{k}}\Big)\Big(1-\exp\big[-i\big(s E_{\mathbf{k}}^{\chi}+E_{\mathbf{p}-\mathbf{k}}^{\phi}\big)t\big]\cos\big(E_{\mathbf{p}}^{\phi}t\big)\Big)\nonumber\\&\qquad-iE_{\mathbf{p}}^{\phi}\exp\big[-i\big(s E_{\mathbf{k}}^{\chi}+E_{\mathbf{p}-\mathbf{k}}^{\phi}\big)t\big]\sin\big(E_{\mathbf{p}}^{\phi}t\big)\Big]\big[1+f\big(sE_{\mathbf{k}}^{\chi}\big)\big]
\nonumber
\\
&
 \qquad +\rho(\mathbf{p}-\mathbf{k},\mathbf{p}'-\mathbf{k};t) 
\,\frac{1}{2E^{\chi}_{\mathbf{k}}2E^{\phi}_{\mathbf{p}-\mathbf{k}} 2E^{\phi}_{\mathbf{p}'-\mathbf{k}}}\,\frac{s}{sE_{\mathbf{k}}^{\chi}+E_{\mathbf{p}-\mathbf{k}}^{\phi}-E_{\mathbf{p}}^{\phi}}\nonumber\\&\qquad\times\Big(1-\exp\big[i(sE_{\mathbf{k}}^{\chi}+E_{\mathbf{p}-\mathbf{k}}^{\phi}-E_{\mathbf{p}}^{\phi}\big)t\big]\Big)f\big(sE_{\mathbf{k}}^{\chi}\big)\Bigg]\: -\:\big(\mathbf{p}\longleftrightarrow \mathbf{p}'\big)^*\Bigg\}
\nonumber\\&\phantom{=\ } -\,\frac{4m^4}{\mathcal{M}^2}\,\rho(\mathbf{p},\mathbf{p}';t)\sum_{s\,=\,\pm}\int_{\mathbf{x}}\int_{\mathbf{k}_1\mathbf{k}_2}\frac{\sin\big[\big(E_{\mathbf{k}_1}^{\phi}+E_{\mathbf{k}_1-\mathbf{k}_2}^{\phi}+sE_{\mathbf{k}_2}^{\chi}\big)t\big]}{E_{\mathbf{k}_1}^{\phi}+E_{\mathbf{k}_1-\mathbf{k}_2}^{\phi}+sE_{\mathbf{k}_2}^{\chi}}\,\frac{s\big[1+f\big(sE_{\mathbf{k}_2}^{\chi}\big)\big]}{2E_{\mathbf{k}_1}^{\phi}2E_{\mathbf{k}_1-\mathbf{k}_2}^{\phi}2E_{\mathbf{k}_2}^{\chi}}\;.
\end{align}
Herein, $\Delta_{xy}^{\rm F}$ and $D_{xy}^{\rm F}$ are the Feynman propagators of the chameleon and `atom' fields, respectively, and \smash{$E^{\chi}_{\mathbf{p}}=\sqrt{\mathbf{p}^2+M^2}$} and \smash{$E^{\phi}_{\mathbf{p}}=\sqrt{\mathbf{p}^2+m^2}$} are their on-shell energies. We have also accounted for thermal fluctuations in the chameleon field (which is assumed to be in thermal equilibrium with the walls of the vacuum chamber), giving rise to the factors of the Bose-Einstein distribution $f(E)=1/(e^{\beta E}-1)$, where $\beta=1/T$ is the inverse thermodynamic temperature (see, e.g., reference~\cite{LeBellac}). We recognize the first term on the right-hand side of equation~\eqref{eq:master} as the free-phase evolution, the remaining terms of the first and second lines as tadpole self-energy insertions, and the terms of the third to seventh lines as the contributions from the one-loop bubble diagram.  The final line contains a disconnected vacuum diagram that we omit hereafter; for further details, see reference~\cite{Burrage:2018pyg}. The self-energies of the second to fifth lines contain quadratic and logarithmic divergences. Noticing, however, that all but the first of the self-energy insertions vanish in the limit $t\to 0$, it follows that the contributions from any counterterms must vanish in the same limit. As a result, the renormalization procedure must be modified to reflect the fact that the external switching on and off of the interactions leads to violation of both time-translational invariance and Lorentz invariance, as described in detail in reference~\cite{Burrage:2018pyg}.

Subtracting the divergences at a reference three-momentum $\bar{\mathbf{p}}$ (see reference~\cite{Burrage:2018pyg}), the master equation can be written in the schematic form
\begin{align}
	\label{eq:masterRen}
	\partial_t\rho(\mathbf{p},\mathbf{p}';t) \: &=\: -\:\big[ i u({\bf p},{\bf p}';t) + \Gamma({\bf p},{\bf p}';t) \big] \rho(\mathbf{p},\mathbf{p}';t) \:+\int_{\bf k} \gamma({\bf p},{\bf p}',{\bf k};t) \rho({\bf p-k},{\bf p'-k};t)\;,
\end{align}
where the various terms are given as follows:
\begin{subequations}
	\label{eq:coeffs}
	\begin{align}
		&u({\bf p},{\bf p}';t) \ \equiv\ E^{\phi}_{\mathbf{p}}\:-\:E^{\phi}_{\mathbf{p}'}\:+\:\frac{m^2}{\mathcal{M}}\left(\frac{1}{E^{\phi}_{\mathbf{p}}}-\frac{1}{E^{\phi}_{\mathbf{p}'}}\right)\Delta^{{\rm F}(T\neq 0)}_{xx}\bigg\{
\frac{1}{\mathcal{M}}
+
\frac{\lambda}{2}\,\braket{X}\,\frac{\cos(Mt)-1}{M^2}\bigg\}\nonumber\\&\ -\:\Bigg\{\frac{1}{E^{\phi}_{\mathbf{p}}}\int_{p^0}\frac{\sin\big[\big(p^0-E^{\phi}_{\mathbf{p}}\big)t\big]}{p^0-E^{\phi}_{\mathbf{p}}}\,{\rm Re}\Big[\,\Pi(-\,p^2)\:-\:\Pi^{(T\,=\,0)}(-\,p^2)\big|_{\mathbf{p}\,=\,\bar{\mathbf{p}}}\Big]\: -\:\big(\mathbf{p}\longleftrightarrow \mathbf{p}'\big)\Bigg\}\;, \label{eq:coeffs_u} \\
		&\Gamma({\bf p},{\bf p}';t) \ \equiv\ \frac{1}{E^{\phi}_{\mathbf{p}}}\int_{p^0}\frac{\sin\big[\big(p^0-E^{\phi}_{\mathbf{p}}\big)t\big]}{p^0-E^{\phi}_{\mathbf{p}}}\,{\rm Im}\,\Pi(-\,p^2)\: +\:\big(\mathbf{p}\longleftrightarrow \mathbf{p}'\big)\;, \label{eq:coeffs_Gamma}\\
		&\gamma({\bf p},{\bf p}',{\bf k};t) \ \equiv \
i\,\frac{4m^4}{\mathcal{M}^2}\sum_{s\,=\,\pm}\Bigg\{
\frac{1}{2E^{\chi}_{\mathbf{k}}2E^{\phi}_{\mathbf{p}-\mathbf{k}} 2E^{\phi}_{\mathbf{p}'-\mathbf{k}}}\,\frac{s}{sE_{\mathbf{k}}^{\chi}+E_{\mathbf{p}-\mathbf{k}}^{\phi}-E_{\mathbf{p}}^{\phi}}\nonumber\\&\phantom{\gamma({\bf p},{\bf p}',{\bf k};t) \ \equiv \ } \times\Big(1-\exp\big[i(sE_{\mathbf{k}}^{\chi}+E_{\mathbf{p}-\mathbf{k}}^{\phi}-E_{\mathbf{p}}^{\phi}\big)t\big]\Big)f\big(sE_{\mathbf{k}}^{\chi}\big) \:-\: ({\bf p} \longleftrightarrow {\bf p}')^*\Bigg\} \;, \label{eq:coeffs_gamma}
	\end{align}
\end{subequations}
where the $t$-dependence should be compared with the diagrammatics of references~\cite{Millington:2012pf,Millington:2013isa} and
\begin{equation}
i\Pi(-\,p^2)\ =\  \bigg(-\frac{2im^2}{\mathcal{M}}\bigg)^2\int_k\bigg[\frac{-\,i}{k^2+M^2-i\epsilon}\:+\:2\pi f(|k^0|)\delta(k^2+M^2)\bigg]\,\frac{-\,i}{(k-p)^2+m^2-i\epsilon}\;.
\end{equation}
The superscripts $T=0$ and $T\neq 0$ indicate respectively the zero-temperature and finite-temperature parts of the various propagators and self-energies. The coefficient $u\in \mathbb{R}$ describes coherent evolution, and $\Gamma\in \mathbb{R}$ and $\gamma\in\mathbb{C}$ describe decoherence and momentum diffusion.

In order to provide a conservative estimate on the order of magnitude of the effects (see reference~\cite{Burrage:2018pyg}), we consider the phase shift that results from the change in the effective mass $m$ of the `atom' due to the ambient value of the chameleon field [cf.~equation~\eqref{eq:massredef}]. (In order to be sensitive to the difference $m-\tilde{m}$, we might imagine a setup in which the ambient value of the chameleon field is different along the two paths of the interferometer.) For a spherical vacuum chamber of radius $1\ {\rm m}$, the value of the chameleon field at the centre of the chamber is $\braket{X}=-1.287/\sqrt{\lambda L^2}$~\cite{Burrage:2014oza, Khoury:2003rn}. Assuming the thermal effects are subleading, and taking $\lambda=1/10$, $\mathcal{M}=M_{\rm Pl}$ (the reduced Planck mass), $\tilde{m}=87m_u$ (the mass of a ${}^{87}{\rm Rb}$ atom in atomic mass units) and a characteristic velocity scale of the atom in the atom interferometer of $v=10\ {\rm ms}^{-1}$, we find the phase shift
\begin{equation}
	|\Delta u| \: \sim\: \big|\braket{X}\tilde{m}v^2/\big(\mathcal{M}c\big)\big|\: \sim\: 10^{-23}\ {\rm Hz}\;,
\end{equation}
which is far below the current sensitivity $\sim 10^{-8}\ {\rm Hz}$ (as inferred from reference~\cite{Estey_2015}).

\section{Concluding remarks}

We have reported a novel way of deriving the quantum master equation for matrix elements of the reduced density operator from first principles in quantum field theory, as presented in reference~\cite{Burrage:2018pyg}, which yields cut-off independent and quantitative predictions that are crucial for ongoing precision searches for dark matter and dark energy. While the context presented here has been screened scalar fifth forces, our approach is general, and its extension to higher-spin fields may find use, for example, in studies of gravitational decoherence.

\ack

This work was supported by a Royal Society University Research Fellowship [CB], a Leverhulme Trust Research Leadership Award [CB and PM], the University of Nottingham Vice-Chancellor's Scholarship for Research Excellence [CK], the UK Science and Technology Facilities Council (STFC) grant no.~ST/L000393/1 [PM], the UK Engineering and Physical Sciences Research Council (EPSRC) grant no.~EP/P026133/1 [JM] and a European Research Council (ERC) grant, agreement no.~335266 (ESCQUMA) [JM].

\end{document}